\documentstyle[12pt]{article}
\title{Some eigenstates for a model associated with solutions
       of tetrahedron equation.\\
       IV.~String---particle marriage}
\author{I.G.~Korepanov\\
\footnotesize $\matrix{ \cr
       \hbox{Chelyabinsk University of Technology}\cr
       \hbox{76 Lenin av., Chelyabinsk 454080, Russia}
       }$}
\date{April 1997}
\def\be{\begin{equation}}
\def\ee{\end{equation}}
\makeatletter
\long\def\@makecaption#1#2{\vskip 10\p@ \hbox to\hsize{\hfil#1\hfil}}
\makeatother
\begin{document}
\maketitle

\begin{abstract}
This paper continues the series begun with works solv-int/9701016,
solv-int/9702004 and solv-int/9703010. Here we construct more
sophisticated strings, combining ideas from those papers
and some considerations involving solutions of tetrahedron equation
due to Sergeev, Mangazeev and Stroganov.
\end{abstract}

\section*{Introduction}

In this work we continue the study, begun in~\cite{I,II,III}, of eigenstates
of the model based on solutions to the tetrahedron equation.
In the paper~\cite{III}, we have been using a kagome lattice made up
of five-legged `$R$-operators' for string construction.
The idea of the present paper
is that probably we will be able to achieve more using
a kagome lattice made up of six-legged `$S$-operators' instead.

Here we will be considering
the infinite in all plane directions kagome lattice.
Let us imagine it as situated in a horizontal plane, having
an $S$-operator in each its vertex and using four of six $S$-operator
legs as lattice edges.
Two other legs of an $S$-operators are vertical.
The $S$-operators form in such way some
{\em auxiliary transfer matrix}.
We will apply some vector---namely, a ``one-particle''
vector from~\cite{I}---to lower
ends of $S$-operators, while a string will be posed, in a sense,
within the plane, as in~\cite{III}.
The result will be new eigenstates for the same ``hedgehog''
transfer matrix as in papers~\cite{I,II,III}.

Such ``hedgehog'' transfer matrix is made up of $S$-operators that arise
in the ``static limit'' of a more general construction
due to Sergeev, Mangazeev and Stroganov (SMS), see paper~\cite{mss}
(operators are called there `$R$' instead of `$S$').
However, the auxiliary transfer matrix will be made of some
{\em other\/} special case of SMS operators described
in Sections \ref{sec-SMS-spec} and~\ref{sec-gauge} below.

As for the rest of the paper,
in Section~\ref{sec-absence} we show that a particle
in absence of a string just vanishes, while their ``marriage''
produces in Section~\ref{sec-marriage} new nontrivial strings.
Discussion is given in Section~\ref{sec-IV-discussion}.

\section{Tetrahedron equation: a special case}
\label{sec-SMS-spec}

As it is known from the paper~\cite{mss}, the solution of the tetrahedron
equation is parameterized by dihedral angles
$\vartheta_1,\ldots,\vartheta_6$ (of which five are independent)
between four planes in the three-dimensional euclidean space.
The equation looks like
\begin{eqnarray}
S_{123}(\vartheta_1, \vartheta_2, \vartheta_3)\,
S_{145}(\vartheta_1, \vartheta_4, \vartheta_5)\,
S_{246}(\pi-\vartheta_2, \vartheta_4, \vartheta_6)\,
S_{356}(\vartheta_3, \pi-\vartheta_5, \vartheta_6)
\nonumber\\
= S_{356}(\vartheta_3, \pi-\vartheta_5, \vartheta_6)\,
S_{246}(\pi-\vartheta_2, \vartheta_4, \vartheta_6)\,
S_{145}(\vartheta_1, \vartheta_4, \vartheta_5)\,
S_{123}(\vartheta_1, \vartheta_2, \vartheta_3) ,
\label{IV-*}
\end{eqnarray}
where the subscripts of $S$ number the spaces where that operator acts.

Let us imagine that those four planes pass through the center
of a unit sphere. Then an alternative way of parameterizing,
say, the operator $S_{123}$ is by using the edges $a_1$, $a_2$
and $a_3$ of the spherical triangle instead of 
$\vartheta_1$, $\vartheta_2$ and $\vartheta_3$. Each vertex
of the triangle is, naturally, the intersection point of the
sphere and a pair of planes belonging to the operator $S_{123}$.

We will assume that the operator $S_{123}$ in the equation~(\ref{IV-*})
is one of the hedgehogs of the ``hedgehog transfer matrix'',
while the other $S$-operators belong to the auxiliary
kagome transfer matrix. Recall that we are considering the hedgehogs
corresponding to the {\em static limit\/} in terms
of~\cite{mss}. This means that the sides $a_1$, $a_2$ and $a_3$
of the spherical triangle are infinitely small.
Let this infinitely small spherical triangle be situated
at the north pole of the sphere.

Now let us fix a particular place for the fourth plane (that doesn't
pass through the north pole). Namely, let this plane
{\em intersect the sphere through its
equator}. The resulting restrictions on the angles $\vartheta$
are the following:
\be
\vartheta_1 +\vartheta_2 +\vartheta_3 =\pi, \qquad
\vartheta_4 =\vartheta_5 =\vartheta_6 ={\pi \over 2}.
\label{IV-**}
\ee
We will see that (\ref{IV-**}) yields the degenerate
operators $S_{145}$, $S_{246}$ and $S_{356}$,
but this only helps to perform the construction.

\section{The gauge for ``doubly rectangular'' $S$-opera\-tors}
\label{sec-gauge}

The explicit form for operators entering in equation~(\ref{IV-*})
can be found in section~5 of the work~\cite{mss}.
In the same work, quite a bit is said about different gauges
for those operators. We will work with the following gauge
for our ``doubly rectangular''---i.e.\ having
two of three angles~$\vartheta$ equal to $\pi/2$---operators:
first take them as in section~5 of~\cite{mss}, and then change
as described in the following paragraph.

Consider e.g.\ the operator $S_{145}=S_{145} (\vartheta_1,
\vartheta_4,\vartheta_5)=S_{145} (\vartheta_1, \pi/2, \pi/2)$.
Its legs corresponding to the spaces number 4 and~5 lie on the
edges of the kagome lattice, in the horizontal plane (see
Introduction). Let us perform a conjugation in both those spaces
with the matrix $\pmatrix{1&-1\cr 1&1}$, that is, let us change
$$
S_{145}\to {\bf 1}\otimes \pmatrix{1&-1\cr 1&1} \otimes
\pmatrix{1&-1\cr 1&1} \cdot S_{145} \cdot {\bf 1}\otimes
\pmatrix{1&-1\cr 1&1}^{-1} \otimes \pmatrix{1&-1\cr 1&1}^{-1}.
$$

Let us perform the similar transformations for $S_{246}$
in the spaces 4 and~6, and for $S_{356}$ in the spaces 5 and~6.
Note that such gauges are consistent with the tetrahedron
equation~(\ref{IV-*}), and that the gauge of $S_{123}$ does not
change.

Now let us forget about the old gauge of the operators
$S_{145}$, $S_{246}$ and $S_{356}$, and use these notations
for the operators in the {\em new\/} gauge. The explicit form
of each of these operators is like this:
$$
S_{\ldots}= {1-\tan(\vartheta/4)\over \cos(\vartheta/2)}
\pmatrix{A&B\cr C&D},
$$
where
\be
A=\pmatrix{
\cos(\vartheta/2) &0&0&0 \cr
0& \sin (\vartheta/2) & 1 &0 \cr
0& 1 & \sin (\vartheta/2) &0 \cr
0&0&0& \cos (\vartheta/2)
},
\label{IV-A}
\ee
\be
B= {\root\of{\sin\vartheta}\over 2} \pmatrix{
1+i &0&0&0 \cr
0& 1-i &0&0 \cr
0&0& -1+i &0 \cr
0&0&0& -1-i
},
\label{IV-B}
\ee
\be
C= {\root\of{\sin\vartheta}\over 2} \pmatrix{
1-i &0&0&0 \cr
0& 1+i &0&0 \cr
0&0& -1-i &0 \cr
0&0&0& -1+i 
},
\label{IV-C}
\ee
\be
D=\pmatrix{
\sin(\vartheta/2) &0&0& -1 \cr
0& \cos(\vartheta/2) &0&0 \cr
0&0& \cos(\vartheta/2) &0 \cr
-1 &0&0& \sin(\vartheta/2)
},
\label{IV-D}
\ee
where
$$
\matrix{
\vartheta= \vartheta_1 \quad \hbox{for } S_{145}, \cr
\vartheta= \vartheta_2 \quad \hbox{for } S_{246}, \cr
\vartheta= \vartheta_3 \quad \hbox{for } S_{356}. }
$$

Matrix $A$ is the matrix of weights for all configurations
of horizontal spins of an $S$-operator, if the spins at the vertical
edges are both fixed and equal~0. Similarly, matrix~$B$
corresponds to the lower spin 1 and the upper spin~0;
matrix~$C$---to the lower spin 0 and the upper spin~1;
and matrix~$D$---to the lower spin 1 and the upper spin~1.

We see that all of these matrices except~$D$ {\em conserve the
total spin}, or, in the other terminology, the ``number of particles'',
within the horizontal plane.

\section{Disappearance of particle in absence of a string}
\label{sec-absence}

Now let us apply the auxiliary kagome transfer matrix made up
of the operators $S_{145}$, $S_{246}$ and $S_{356}$, to the
one-particle vector from the papers~\cite{I,II}.
We will choose the following boundary conditions at the horizontal
infinity: all spins at the ``far enough'' horizontal edges
are zero.

Let us show that in this case the spin at {\em all\/} horizontal edges
are zero. Recall that the one-particle state is a linear combination
of tensor products of spins, of which {\em one\/} equals unity,
and the others are zero. This means that matrix~$D$ (see the end of
section~\ref{sec-gauge})---the only one that can change the total
spin---cannot appear twice. In other words, non-zero spin cannot
be created somewhere and then annihilated somewhere else,
thus all the horizontal edges possess zero spins.

This conclusion implies that only {\em upper left\/} entries
of the matrices $A,\ldots,D$ are playing r\^ole. For a given
$S$-operator, four such elements form a matrix
\be
{1-\tan(\vartheta/4)\over \cos(\vartheta/2)} \pmatrix{
\cos(\vartheta/2) & {\textstyle 1+i\over \textstyle 2}
\,\root\of{\sin\vartheta} \cr
\noalign{\smallskip}
{\textstyle 1-i\over \textstyle 2}
\,\root\of{\sin\vartheta} & \sin(\vartheta/2)
},
\label{IV-deg}
\ee
that transforms a two-dimensional vector at the lower vertical edge
to the one at the upper vertical edge, and each $S$-operator of the kagome
lattice does this independently.

The remarkable property of the matrix
(\ref{IV-deg}) is its {\em degeneracy}. Thus, the whole lattice
transfer matrix applied to the one-particle state yields
an {\em infinite sum of vectors proportional to a fixed vector}.
This sum must be inevitably equal to zero, due to the infiniteness
of lattice in all directions, the fact that a one-particle vector
just acquires some scalar factors under lattice translations,
and a logical assumption
$$
\sum_{n=-\infty}^{\infty} a^n =0
$$
for the sum of an infinite in both directions geometrical progression.

\section{The marriage: an example of a string}
\label{sec-marriage}

The matrix (\ref{IV-deg}) is not the only degenerate one.
Formulae~(\ref{IV-A}--\ref{IV-D}) show really that if we fix
{\em any\/} values (0 or 1) for the four horizontal spins
of an $S$-operator, then the $2\times 2$-matrix corresponding to the
two remaining vertical edges is degenerate. Imagine now that we have fixed
the spins at all horizontal edges of {\em all operators}. Then
edges with the spin~1 form ``strings'', and a remarkable conclusion
is that a configuration of those strings determines the ``outgoing''
vector at the upper vertical edges up to a scalar factor.

\medbreak

{\bf Remark. }However, the ``incoming'' vector at the lower edges
can cause this scalar factor to equal~0 for some string configurations.
For example, for a one-particle incoming vector, there can be no
closed strins, because they require at least two matrices~$D$
to be involved, as explained in Section~\ref{sec-absence}.

\medbreak

We can propose the following
example of a string-like state resulting from the ``marriage''
of a one-particle state and a string within the horizontal plane.
The one-particle state is applied to the lower edges, as already
explained. It is supposed that the string is ``born'' in the vertex
where the particle is applied, due to the matrix~$D$ which is
allowed to change by~2 the total ``horizontal'' spin.
The form of the string at infinity let be fixed by two given
half-lines, each going along the edges of the lattice
in one of the east, north, or north-east directions.

\section{Discussion}
\label{sec-IV-discussion}

In this paper, we have presented the simplest nontrivial states
of the model on infinite lattice that are, in essence,
strings from~\cite{III} fertilized by nontrivial momentum particles
of~\cite{I,II}. The main point is that such string, in contrast with
the ones from~\cite{III}, are not just invariant under the shifts
along themselves, but acquire some nontrivial multipliers.

It is clear that really a quite immense zoo
of such states can be constructed.
It seems also that a string from Section~\ref{sec-marriage}
can be made {\em closed\/} and put
in a {\em finite\/} lattice.

 From the purely mathematical point of view, our construction
of some species of the zoo of $2+1$-dimensional model eigenstates
can be regarded a modest step from the $1+1$-dimensional quantum groups
to the studying of higher-dimensional integrability.

\end{document}